\documentclass[amsmath,amssymb,aps,nofootinbib,prd,reprint,superscriptaddress]{revtex4-2}

\usepackage[T1]{fontenc}
\usepackage[dvipsnames]{xcolor}
\usepackage{hyperref}
\hypersetup{colorlinks=true,citecolor=RoyalBlue,linkcolor=RoyalBlue,urlcolor=RoyalBlue}

\let\originalleft\left
\let\originalright\right
\renewcommand{\left}{\mathopen{}\mathclose\bgroup\originalleft}
\renewcommand{\right}{\aftergroup\egroup\originalright}
\mathcode`\*="8000
{\catcode`\*=\active\gdef*{\mathclose{}\,\mathopen{}}}

\newcommand{\ab}[1]{\left|#1\right|}
\newcommand{\av}[1]{\left\langle#1\right\rangle}
\newcommand{\br}[1]{\left[#1\right]}
\newcommand{\cu}[1]{\left\{#1\right\}}
\newcommand{\pa}[1]{\left(#1\right)}
\newcommand{\ed}{\mathop{}\!\mathrm{d}}
\newcommand{\pd}{\mathop{}\!\partial}
\newcommand{\SD}{\mathcal{R}_1}
\newcommand{\sg}{\operatorname{sg}}
\newcommand{\tlam}{\tilde{\lambda}}

\begin{document}

\title{Single-minus gluon tree amplitudes are nonzero}

\author{
Alfredo Guevara,\textsuperscript{1}
Alexandru Lupsasca,\textsuperscript{2,3}
David Skinner,\textsuperscript{4}\\
Andrew Strominger,\textsuperscript{5}
and Kevin Weil\textsuperscript{2}
on behalf of OpenAI\\
\scriptsize
\textsuperscript{1}Institute for Advanced Study\quad
\textsuperscript{2}OpenAI\quad
\textsuperscript{3}Vanderbilt University\quad
\textsuperscript{4}Cambridge University\quad
\textsuperscript{5}Harvard University}
\noaffiliation

\begin{abstract}
Single-minus tree-level $n$-gluon scattering amplitudes are reconsidered.
Often presumed to vanish, they are shown here to be nonvanishing for certain ``half-collinear'' configurations existing in Klein space or for complexified momenta.
We derive a piecewise-constant closed-form expression for the decay of a single minus-helicity gluon into $n-1$ plus-helicity gluons as a function of their momenta.
This formula nontrivially satisfies multiple consistency conditions including Weinberg's soft theorem.
\end{abstract}

\maketitle

The laws of physics are succinctly encoded in \emph{scattering amplitudes}, which give the quantum probabilities for any given collection of incoming particles to collide and produce any given collection of outgoing particles.
These amplitudes may be systematically derived from the Feynman diagram expansion, which perturbatively sums over all possible quantum processes.
Theoretical results from the Feynman diagram expansion of the Standard Model agree with experiment to an unprecedented 14 decimal places \cite{Aoyama:2017uqe,Morel:2020dww,Fan:2022eto}.
 
In practice, the computation of scattering amplitudes can be extremely difficult.\footnote{The aforementioned agreement between theory and experiment required over a half century of analytic and numerical work.}
Among other obstacles, the growth in the number of Feynman diagrams for an $n$-particle amplitude is faster than exponential in $n$.
However, despite this apparent complexity, cancellations lead in a variety of contexts to a very simple final answer.
This indicates that our present understanding of the quantum laws of physics is seriously incomplete and that a more efficient formulation is needed.
The last few decades have seen much effort in this direction and delivered promising insights; see, \emph{e.g.},~\cite{Dixon:1996wi,Witten:2003nn,Roiban:2004yf,Bern:2005iz,Britto:2005fq,Arkani-Hamed:2012zlh,Arkani-Hamed:2013jha}.

A prominent example of this phenomenon arises in the tree-level color-ordered scattering of gluons---the particles that mediate the strong force and comprise Yang--Mills theory.
Naively, the $n$-gluon scattering amplitude involves order $n!$ terms.
Famously, for the special case of MHV (maximally helicity violating) tree amplitudes, Parke and Taylor \cite{Parke:1986gb} gave a simple and beautiful, closed-form, single-term expression for all $n$.

By definition, $n$-gluon MHV amplitudes have $2$ minus-helicity particles and $n-2$ plus-helicity gluons, which for generic (complexified) kinematics at tree level is the maximally allowed number \cite{Parke:1986gb,Grisaru:1977px,Elvang:2013cua,Dixon:1996wi,Bern:1994zx}.
This gives them a privileged role in the theory, enabling their use as efficient building blocks for the full Yang--Mills theory.

In general, $n-2$ is actually \emph{not} the maximally allowed number of plus gluons.
In this paper, we show that $n-1$ plus (or ``single-minus'') amplitudes are in fact allowed even at tree level\footnote{Witten~\cite{Witten:2003nn} notes that single-minus tree amplitudes are supported at a point in twistor space; see also~\cite{Roiban:2004yf}.
All-plus and single-minus amplitudes are generically corrected at loop level \cite{Bern:1993qk,Mahlon:1993si}.}
with restricted ``half-collinear'' kinematics.\footnote{The half-collinear condition can be viewed as restricting the ingoing and outgoing momenta to a one-dimensional null circle on the celestial torus at the boundary of Klein space.}
The amplitude is divided into chambers whose walls are regions where sums of various subsets of the half-collinear momenta are orthogonal as described below.
The (stripped) amplitudes are piecewise-constant integers in each chamber.
These values are determined from the perturbative Berends--Giele recursion \cite{Berends:1987me}, which is equivalent to Feynman diagrams.

Moreover, for the special kinematic region corresponding to single-minus decay into $n-1$ plus, we give a simple formula for all $n$.
In this special region, the stripped amplitude only takes the values of $+1$, $-1$, or $0$.

The key formula \eqref{eq:Simplification} for the amplitude in this region was first conjectured by GPT-5.2 Pro and then proved by a  new internal OpenAI model.
The solution was checked by hand using the Berends--Giele recursion and was moreover shown to nontrivially obey the soft theorem, cyclicity, Kleiss--Kuijf, and $\mathsf{U}(1)$ decoupling identities---none of which are evident from direct inspection.

The structural role of these single-minus amplitudes in Yang--Mills theory remains to be understood.
We note that, while our expression is a dramatic simplification of the direct Feynman-diagram expression, it is entirely possible that a yet simpler expression may be obtained with a clever choice of analytic continuation, variables or basis, even outside the single-minus decay channel. 
We suspect that there are more interesting insights to come with our methodology and hope that this paper is a step on the road to a more complete understanding of the inner structure of scattering amplitudes.

Single-minus amplitudes also arise in self-dual Yang--Mills theory (SDYM) \cite{Cangemi:1996rx}, a restricted sector of Yang--Mills, and potentially resolve a puzzle therein.
In general, the tree amplitudes of the Feynman expansion are thought to be equivalent to the fully nonlinear classical theory.
However, on the one hand the classical solution space of SDYM is extremely nontrivial \cite{WardWells:1990twistor,Ward:1977sdym_twistor,ADHM:1978instantons}, while the tree diagrams were previously supposed to yield trivial two-point and three-point expressions.
The latter seem insufficient to reproduce the former.
Potentially, the single-minus tree amplitudes in SDYM found here resolve this tension.

This paper is organized as follows.
In Sec.~\ref{sec:Recursion}, we set up notation, describe the standard MHV amplitudes, explain how half-collinear single-minus amplitudes evade the usual no-go condition, and then derive the general Berends--Giele recursion relation.
The solution passes various consistency checks, including the soft theorem, and we provide explicit formulas up to $n=6$ points, where there are already 32 terms.
In Sec.~\ref{sec:Formula}, we restrict to a special kinematic channel denoted $\SD$ with one ingoing minus and $n-1$ outgoing plus gluons.
There, using various identities through $n=6$, we find the answer can be expressed as a signed product of $n-2$ projection operators.
This motivates a guess for the all-$n$ formula, which we verify directly via the Berends--Giele recursion.
We derive a multi-$\delta$-function identity in App.~\ref{app:Master} and give more details of the single-minus specialization of the Berends--Giele recursion in App.~\ref{app:Derivation}.

Further details of our analysis, including a longer general formula for the single-minus amplitude outside of $\SD$, will appear elsewhere.
Our main result immediately leads to a number of extensions.
The construction generalizes directly from gluon to graviton amplitudes and has a simple supersymmetrization.
The results should transform under the $S$-algebra, the ${\cal L}w_{1+\infty}$ algebra \cite{Guevara:2021abz,Strominger:2021mtt}, and their supersymmetric extensions.
In the context of celestial holography, the Mellin transform of the amplitudes in some sectors is given by Lauricella functions.
These results will be reported elsewhere.

\subsection{Notation and useful identities}

This subsection defines our notation\footnote{Our conventions are close to those of \cite{Witten:2003nn}, except for a factor of 2 on the LHS of (2.7) therein.}
and presents several useful identities. 
We use spinor-helicity variables for massless momenta \cite{Elvang:2013cua}
\begin{align}
    p_{\alpha\dot\alpha}=\lambda_\alpha\tlam_{\dot\alpha},
\end{align}
where $(\lambda,\tlam)$ are \emph{real} spinors in $(2,2)$ Klein signature.
As usual in the description of scattering, it is convenient to fix a suitable Lorentz and helicity frame to perform the calculation, which can be restored at the end.
With the benefit of hindsight, we write
\begin{align}
    \label{eq:Conventions}
    |i\rangle=\lambda_i=\pa{1,z_i},\qquad
    |i]=\tlam_i=\omega_i\pa{1,\tilde{z}_i},
\end{align}
with $z_i$ and $\tilde{z}_i$ real and independent.
We use standard brackets for contracting helicity spinors,
\begin{align}
    \langle ij\rangle&=\av{\lambda_i\lambda_j}
    =\epsilon_{\alpha\beta}\lambda_i^\alpha\lambda_j^\beta,\\
    [ij]&=\br{\tlam_i\tlam_j}
    =\epsilon_{\dot\alpha\dot\beta}\tlam_i^{\dot\alpha}\tlam_j^{\dot\beta},
\end{align}
with $p_{ij}^2=(p_i+p_j)^2=\langle ij\rangle [ij]$.
In our parameterization,
\begin{align}
    \av{ij}=z_{ij},\qquad
    [ij]=\omega_i\omega_j\tilde{z}_{ij},
\end{align}
where $z_{ij}:=z_i-z_j$ and $\tilde{z}_{ij}:=\tilde{z}_i-\tilde{z}_j$.
We take our polarization vectors to be
\begin{align}
    \label{eq:PolarizationVectors}
    \epsilon^-_j=\sqrt{2}\frac{|r]\langle j|}{[rj]},\qquad
    \epsilon^+_k=\sqrt{2}\frac{|k]\langle r|}{\av{rk}},
\end{align}
where $|r\rangle$ and $|r]$ are arbitrary reference spinors.
We note that with our conventions~\eqref{eq:Conventions} for fixing the little group frame, $\epsilon^{\pm}$ has mass dimension $\pm 1$, which will affect the mass dimensions of the amplitudes given below.
To avoid proliferation of factors of $2\pi$, we normalize all $\delta$-functions such that
\begin{align}
    \label{eq:DeltaNormalization}
    \int\delta(x)\ed x=2\pi,\qquad
    \frac{1}{x+i\epsilon}-\frac{1}{x-i\epsilon}\stackrel{\epsilon\to0}{=}-i\,\delta(x).
\end{align}
Throughout this paper, we use the standard Feynman propagator $1/(p^2+i\epsilon)$.
Other prescriptions have been considered in Klein signature~\cite{Arkani-Hamed:2009hub}.

To clarify our conventions, with our normalization, the $n$-point MHV (double-minus) color-ordered tree amplitude $\mathcal{A}_n^{\rm MHV}(1^+,\ldots r^-,\ldots s^-,\dots,n^+)$ is
\begin{align}
    \label{eq:NoEpsilonMHV}
    \mathcal{A}_n^{\rm MHV}=i\frac{\av{rs}^4}{\av{12}\av{23}\cdots\av{n1}}\,\delta^4\pa{\sum_{k=1}^n p_k}.
\end{align}
In fact, for full generality, we will need to be careful about the $i\epsilon$ prescription.
We therefore introduce the regularized Parke-Taylor factor
\begin{align}
    \mathrm{PT}_{\rm cyc}=\prod_{k=1}^{n}\frac{[k,k+1]}{p^2_{k,k+1}+i\epsilon}
    =\prod_{k=1}^{n}\frac{1}{z_{k,k+1}{+}i\epsilon \sg_{k,k+1}},
\end{align}
with $n+1\equiv 1$, and where we have defined
\begin{align}
    \sg_{ij}=\sg\pa{[\tlam_i\tlam_j]}
\end{align}
in the frame~\eqref{eq:Conventions}.\footnote{Without fixing a frame, we would take $\sg_{ij}=\sg\pa{[ij]\av{ir}\av{jr}}$, where $|r\rangle$ is any fixed reference spinor.
In $\mathcal{A}_n^{\rm MHV}$, the difference in choices of $|r\rangle$ can be absorbed by $\epsilon$.}
Here, $\sg(x)=2\Theta(x)-1$ denotes the sign function and $\Theta(x)$ is the step function.
The MHV tree amplitude can then be written using $\mathrm{PT}_{\rm cyc}$ as
\begin{align}
    \label{eq:MHV}
    \mathcal{A}_n^{\rm MHV}=i\av{rs}^4\,\mathrm{PT}_{\rm cyc}\,\delta^4\pa{\sum_{k=1}^n p_k}.
\end{align}
Away from walls where $\av{k\,\,k\!+\!1}=0$, we may ignore the $i\epsilon$ prescription and \eqref{eq:MHV} reduces to~\eqref{eq:NoEpsilonMHV}.

It will also be useful to define an ``incomplete'' or ``open chain'' Parke-Taylor factor as
\begin{align}
    \label{eq:PTC}
    \mathrm{PT}_{1\cdots n}=\prod_{k=1}^{n-1}\frac{[k,k+1]}{p^2_{k,k+1}+i\epsilon}
    =\prod_{k=1}^{n-1}\frac{1}{z_{k,k+1}{+}i\epsilon\sg_{k,k+1}}.
\end{align}
This incomplete factor $\mathrm{PT}_{1\cdots n}$ is what naturally appears in App.~\ref{app:Derivation} inside the Berends--Giele recursion as the denominator of the off-shell current with momentum $p_{1\cdots n}$: the cyclic factor is ``opened'' because one leg is off-shell.

\section{Single-minus amplitudes}
\label{sec:Recursion}

In this section, we first explain why the standard argument that the single-minus $n$-particle tree amplitudes vanish in fact fails when all the external particles become collinear.
We then present a recursion relation, derived in App.~\ref{app:Derivation}, that determines these amplitudes for all $n$. 

\subsection{The half-collinear regime}

The kinematic locus we call the \emph{half-collinear regime} is defined by
\begin{align}
    \av{ij}=0\qquad\forall\,i,j\in\cu{1,\ldots,n}.
\end{align}
In (2,2) signature, this is compatible with nonzero $[ij]$, unlike in Minkowski space.\footnote{The half-collinear regime also makes sense for complex momenta.
We believe single-minus amplitudes also exist with complex momenta; it would be interesting to understand their continuation.}
In the frame \eqref{eq:Conventions}, this locus implies all $z_{ij}=0$ but does not restrict the $\omega_i$ or $\tilde{z}_i$.

We will now show that single-minus tree amplitudes can be nonzero in the half-collinear regime.
This can be demonstrated by exposing the ``loophole'' in the power-counting argument that single-minus amplitudes vanish.
We choose polarization vectors for the $n$ gluons as in \eqref{eq:PolarizationVectors}
\begin{align}
    \label{eq:Polarization}
    \epsilon^-_1=\sqrt{2}\frac{|r]\langle1|}{[r1]},\qquad
    \epsilon^+_a=\sqrt{2}\frac{|r\rangle[a|}{\av{ra}}\qquad
    \text{for}\ a\ge 2,
\end{align}
where $|r]$ and $|r\rangle$ are arbitrary reference spinors.
Now, for generic kinematics, if we choose $|r\rangle=|1\rangle$, then we find that all the polarization vectors are orthogonal.
As such, the amplitude can be nonzero only if they are contracted with momenta in the numerator.
Powers of momenta in the numerator appear only from vertices, and there are at most $n-2$.
These are insufficent to contract with all the polarization vectors.
Hence, single-minus amplitudes vanish for generic kinematics.

The loophole in the argument is that we cannot choose $|r\rangle=|1\rangle$ if $\av{1a}=0$ for any $|a\rangle$, as the polarization vectors $\epsilon^+_a$ would become singular.
Therefore, we cannot conclude that the amplitude vanishes on the locus where $\av{1a}=0$.
In fact, the single-minus 3-point amplitude (also anti-MHV) is known to have a factor $\delta(\av{12})\delta(\av{13})$ restricting to this locus.
Moreover, it can be shown by induction that the $n$-point amplitude may only be supported when \emph{all} $\av{ij}=0$.

To express the fact that it is supported in the half-collinear regime, the single-minus tree-level $n$-gluon amplitude $\mathcal{A}_n(1^-,2^+,\dots,n^+)$ can be written as
\begin{align}
    \label{eq:AnsatzP}
    \mathcal{A}_n&=i^{2-n}\frac{\av{r1}^{n+1}}{\av{r2}\av{r3}\cdots\av{rn}}A_{1\cdots n}\prod_{a=2}^n\delta(\av{1a})\notag\\
    &\qquad\times\delta^2\pa{\sum_{i=1}^n\av{ri}\tlam_i}.
\end{align}
Here, we introduced the \emph{stripped amplitude} $A_{1\cdots n}$.
The $\delta$-functions imposing $\av{1a}=0$ simply ensure the full amplitude is supported only in the half-collinear regime, while the $\delta$-functions in $\tlam$ enforce the remaining components of momentum conservation.
The prefactor ensures that $\mathcal{A}_n$ has the correct little-group scaling for a single-minus amplitude.
The collinear $\delta$-functions ensure that this prefactor and the $\tlam$ $\delta$-functions are independent of the reference spinor $|r\rangle$, as long as it is chosen so that $\langle r1\rangle\neq0$.

The interest is in the stripped amplitude $A_{1\cdots n}$, which carries no helicity weight and depends only on kinematics.
In the frame \eqref{eq:Conventions} and picking $|r\rangle=(0,1)$, $A_{1\cdots n}$ is a function only of $\{\tilde{\lambda}_i\}$, and \eqref{eq:AnsatzP} becomes\footnote{Some readers may find more intuitive an alternative expression, obtained in the frame $\tilde{\lambda}\to\frac{1}{\sqrt{|\omega|}}\tilde{\lambda}$, $\lambda\to\sqrt{|\omega|}\lambda$, which renders the polarizations \eqref{eq:Polarization} dimensionless and gives $\mathcal{A}_n$ scaling as $\omega^{-n}$:
\begin{align*}
    \mathcal{A}_n\to i^{2-n}\ab{\frac{\omega_1}{\omega_2\omega_3\cdots\omega_n}}A_{1\cdots n}\prod_{a=2}^n\delta(z_{1a})\,\delta^2\pa{\sum_{i=1}^n\sqrt{|\omega_i|}\tlam_i}.
\end{align*}}
\begin{align}
    \label{eq:Ansatz}
    \mathcal{A}_n=i^{2-n}A_{1\cdots n}\prod_{a=2}^n\delta(z_{1a})\,\delta^2\pa{\sum_{i=1}^n\tlam_i}.
\end{align}
We will sometimes use the shorthand
\begin{align}
    \delta_{1\cdots n}=i^{1-n}\prod_{k=1}^{n-1}\delta(z_{k,k+1})
\end{align}
to denote these half-collinear $\delta$-functions.

\subsection{The recursion relation}

The first main result of this paper is the recursion relation presented in~\eqref{eq:Recursion} below.
This relation determines all $n$-particle single-minus tree amplitudes.
Solving it is equivalent to, but slightly simpler than, summing the Feynman diagrams for these amplitudes. 

For any ordered list $S=(q,\ldots,p)$, we first define the list momentum $\tlam_S=\sum_{i\in S}\tilde{\lambda}_i$ using the frame~\eqref{eq:Conventions}.\footnote{In a generic frame, $\tlam_S=\sum_{i\in S}\tlam_i\av{ir}$ for any reference spinor $|r\rangle$.
The dependence of both this expression and the sign functions on $|r\rangle$ drops out on the support of the collinear $\delta$-functions.}
We then define the \emph{preamplitude} $\bar A_S$ by taking
\begin{align}
    \bar{A}_{q}=1,\qquad
    \bar{A}_{qp}=0,
\end{align}
when $|S|=1$ and $|S|=2$, and extending recursively to $|S|\geq3$ via
\begin{align}
    \label{eq:Preamplitudes}
    \bar{A}_{q\cdots p}=-\sum_{\,\text{o.p.}}V_{\tilde\lambda_{S_1}\cdots\tilde\lambda_{S_A}}\;\prod_{a=1}^{A}\bar{A}_{S_a},
\end{align}
where the sum is over all ordered partitions of $(q\cdots p)=(S_1|S_2|\cdots|S_A)$ into $A\geq3$ parts.

Here, the vertex term $V_{\tilde\lambda_{S_1}\cdots\tilde\lambda_{S_k}}$ is defined by  
\begin{align}
    \label{eq:Vertex}
    V_{\tlam_1\cdots\tlam_n}&=\prod_{k=1}^{n-1}\sg_{k,k+1}\Theta\pa{-\frac{\br{\tlam_{1\cdots k}\tlam_{k+1\cdots n}}}{\br{\tlam_k\tlam_{k+1}}}},
\end{align}
in the case where each block $S_a$ contains only one element.
We also take $V_{\tlam_1}=1$.
Having determined $\bar A_S$, the stripped amplitude $A_{1\cdots n}$ itself is then given by 
\begin{align}
    \label{eq:Recursion}
    A_{1\cdots n}=-\sum_{\,\text{o.p.}}\widehat{\mathrm{PT}}_{\tlam_{S_1}\cdots\tlam_{S_A}}\;\prod_{a=1}^{A}\bar A_{S_a},
\end{align}
where the ordered partition of $(2\ldots n)$ now has $A\geq1$ parts, while
\begin{align}
    \widehat{\mathrm{PT}}_{\tlam_1\cdots\tlam_n}=V_{\tlam_1\cdots\tlam_n}-\bar V_{\tlam_1\cdots\tlam_n},
\end{align}
where $\bar V_{\tlam_1\cdots\tlam_n}$ is given by \eqref{eq:Vertex} with a $+$ sign within the argument of $\Theta$.
We may refer to this object as \emph{on-shell Parke--Taylor}, as it is related to $\rm PT_{1\cdots n}$ by a standard LSZ reduction procedure fleshed out in App.~\ref{app:Derivation}.

It is also useful to note that the incomplete Parke-Taylor factor $\mathrm{PT}_{1\cdots n}$, the collinear $\delta$-functions and the vertex function $V_{\tlam_1\cdots\tlam_n}$ are related via the useful identity
\begin{align}
    \label{eq:PT}
    &\mathrm{PT}_{1\cdots n}-\delta_{1\cdots n}\,V_{\tlam_1\cdots\tlam_n}\notag\\
    &=\sum_{j=1}^{n-1}\frac{\br{\tlam_{1\cdots j}\tlam_{j+1\cdots n}}}{p_{1\cdots n}^2+i\epsilon}\mathrm{PT}_{1\cdots j}\,\mathrm{PT}_{j+1\cdots n},
\end{align}
which follows from a master identity given in App.~\ref{app:Master}.

\subsection{Consistency checks}
\label{sec:Properties}

It follows from the definition \eqref{eq:AnsatzP} that the stripped amplitudes $A_{12\cdots n}$ satisfy the following properties:
\begin{enumerate}
    \item Cyclicity:
    \begin{align}
    \label{eq:Cyclicity}
        A_{12\cdots n}=A_{2\cdots n1}.
    \end{align}
    \item Reflection symmetry:
    \begin{align}
        A_{12\cdots n}=(-1)^nA_{n\cdots 21}.
    \end{align}
    \item $\mathsf{U}(1)$ decoupling:
    \begin{align}
        A_{12\cdots n}+A_{13\cdots n2}+A_{14\cdots 23}+\ldots=0.
    \end{align}
    \item Kleiss--Kuijf (KK) relations.
    For instance, at $n=5$,
    \begin{align}
        A_{12345}+A_{12354}+A_{12435}+A_{14235}=0.
    \end{align}
    (The general form of these relations can be found, \emph{e.g.}, in~\cite{Elvang:2013cua}).
    \item Weinberg's soft theorem:\footref{SoftTheoremFootnote}
    \begin{align}\label{eq:softm}
        \lim_{\omega_n\to 0}A_{1\cdots n}=\frac{1}{2}\pa{\sg_{n-1,n}+\sg_{n1}}A_{1\cdots n-1}\,.
    \end{align}
\end{enumerate}

It is far from evident that all these properties are obeyed by the solution of the recursion relation \eqref{eq:Recursion}.
Nonetheless, we have verified by explicit calculation that they do indeed hold.
Details of this calculation will appear elsewhere.

\subsection{Concrete examples}

From \eqref{eq:Recursion}, the 3-point to  6-point single-minus stripped amplitudes are, using  $\sg_{i,jk}=\sg\pa{\br{\tlam_i,\tlam_j+\tlam_k}}$, \emph{etc.},
\begin{align}
    A_{123}&=\sg_{12},\\
    A_{1234}&=\frac{1}{2}\pa{\sg_{23}\sg_{41}+\sg_{12}\sg_{34}};\\
    A_{12345}=\frac{1}{4}\Big[&\sg_{51}\sg_{34}\sg_{2,34}+\sg_{51}\sg_{23}\sg_{23,4}\\
    &-\sg_{51}\sg_{2,34}\sg_{23,4}+\sg_{45}\sg_{23}\sg_{1,23}\notag\\
    &+\sg_{45}\sg_{12}\sg_{12,3}-\sg_{45}\sg_{1,23}\sg_{12,3}\notag\\
    &+\sg_{51}\sg_{45}\sg_{12,34}+\sg_{12}\sg_{34}\sg_{12,34}\Big]\notag,
\end{align}
\begin{widetext}
\begin{align}
    \label{eq:A123456}
    A_{123456}=\frac{1}{8}\Big[
        &-\sg_{1,23}\sg_{12,3}\sg_{123,4}\sg_{56}
        +\sg_{1,23}\sg_{123,4}\sg_{23}\sg_{56}
        +\sg_{1,234}\sg_{12,34}\sg_{123,4}\sg_{56}
        -\sg_{1,234}\sg_{12,34}\sg_{34}\sg_{56}\notag\\
        &-\sg_{1,234}\sg_{123,4}\sg_{23}\sg_{56}
        -\sg_{1,234}\sg_{2,34}\sg_{23,4}\sg_{56}
        +\sg_{1,234}\sg_{2,34}\sg_{34}\sg_{56}
        +\sg_{1,234}\sg_{23}\sg_{23,4}\sg_{56}\notag\\
        &+\sg_{12}\sg_{12,3}\sg_{123,4}\sg_{56}
        -\sg_{12}\sg_{12,34}\sg_{123,4}\sg_{56}
        +\sg_{12}\sg_{12,34}\sg_{34}\sg_{56}
        +\sg_{12}\sg_{345,6}\sg_{45,6}\sg_{56}\notag\\
        &-\sg_{1,23}\sg_{12,3}\sg_{45}\sg_{45,6}
        +\sg_{1,23}\sg_{23}\sg_{45}\sg_{45,6}
        +\sg_{12}\sg_{12,3}\sg_{45}\sg_{45,6}
        -\sg_{12}\sg_{3,45}\sg_{34,5}\sg_{345,6}\notag\\
        &+\sg_{12}\sg_{3,45}\sg_{345,6}\sg_{45}
        +\sg_{12}\sg_{34}\sg_{34,5}\sg_{345,6}
        -\sg_{2,34}\sg_{23,4}\sg_{234,5}\sg_{61}
        +\sg_{2,34}\sg_{234,5}\sg_{34}\sg_{61}\notag\\
        &+\sg_{2,345}\sg_{23,45}\sg_{234,5}\sg_{61}
        -\sg_{2,345}\sg_{23,45}\sg_{45}\sg_{61}
        -\sg_{2,345}\sg_{234,5}\sg_{34}\sg_{61}
        -\sg_{2,345}\sg_{3,45}\sg_{34,5}\sg_{61}\notag\\
        &+\sg_{2,345}\sg_{3,45}\sg_{45}\sg_{61}
        +\sg_{2,345}\sg_{34}\sg_{34,5}\sg_{61}
        +\sg_{23}\sg_{23,4}\sg_{234,5}\sg_{61}
        -\sg_{23}\sg_{23,45}\sg_{234,5}\sg_{61}\notag\\
        &+\sg_{23}\sg_{23,45}\sg_{45}\sg_{61}
        +\sg_{345,6}\sg_{45}\sg_{45,6}\sg_{61}
        +\sg_{23}\sg_{45,6}\sg_{56}\sg_{61}
        +\sg_{34}\sg_{345,6}\sg_{56}\sg_{61}\Big].
\end{align}
\end{widetext}
\footnotetext{\label{SoftTheoremFootnote}Through \eqref{eq:Ansatz}, this is seen to follow from the conventional form $\mathcal{A}_{n}\to\pa{\frac{\av{1r}}{\av{nr}}\frac{[n,1]}{p_{n,1}^2+i\epsilon}-\frac{\av{n-1,r}}{\av{nr}}\frac{[n,n-1]}{p_{n,n-1}^2+i\epsilon}}\mathcal{A}_{n-1}$.}
Clearly a more concise formula is needed!

\section{Amplitudes in the first region}
\label{sec:Formula}

This section presents the next main result of this paper: a simple formula for the $n$-point single-minus amplitudes \eqref{eq:Recursion} with partially restricted kinematics within the half-collinear regime.

\subsection{Restricted kinematics within the half-collinear regime}

We define the kinematic region $\SD$ by the condition that, with $\tlam_{\dot\alpha}=\omega(1,\tilde{z})$, there exists at least one $\mathsf{SO}(2,2)$ frame in which
\begin{align}
    \label{eq:Pn}
    \omega_1<0,\qquad
    \omega_a>0,\qquad
    a\in\cu{2,\ldots,n}.
\end{align}
This region is fully consistent with the half-collinear regime where all $\av{ij}=0$ and further restricts the kinematics within that regime.
Unlike in Minkowski signature, we note that in Klein signature, there is no invariant meaning to a particle having positive frequency.
Nonetheless, $\SD$ is $\mathsf{SO}(2,2)$-invariant because we only ask that there exists \emph{some} frame in which~\eqref{eq:Pn} holds.
Geometrically, it amounts to requiring that there exists some straight line through the origin in $\mathbb{R}^2$ such that $\tlam_{1}$ lies on one side and all other $\tlam_a$ lie on the other.
$\SD$ describes a single ingoing self-dual gluon decaying to $n-1$ outgoing anti-self-dual gluons, where ingoing/outgoing refers to the frame in which the inequalities~\eqref{eq:Pn} hold.

Interestingly, we will see that the amplitudes dramatically simplify in $\SD$, where certain sign functions become independent of the frequencies $\omega_k$.
In particular,
\begin{align}
    \label{eq:SignSimplification}
    \sg_{ij}=\sg\tilde{z}_{ij},\qquad
    \sg_{1j}=\sg\tilde{z}_{j1}\qquad
    \forall\,i,j\ge2.
\end{align}
Note however that the $\omega_k$ cannot be eliminated from expressions such as $\sg_{2,34}$.

\subsection{Concrete examples}

In region $\SD$, using \eqref{eq:SignSimplification}, momentum conservation and spinor identities, one may show that the long expressions of the previous section dramatically simplify to 
\begin{align}
    \label{eq:PTT3}
    A_{123}|_{\SD}&=\frac{1}{2}\pa{\sg_{12}+\sg_{23}},\\
    A_{1234}|_{\SD}&=\frac{1}{4}\pa{\sg_{12}+\sg_{23}}\pa{\sg_{34}+\sg_{41}},\\
    A_{12345}|_{\SD}&=\frac{1}{8}\pa{\sg_{12}+\sg_{23}}\pa{\sg_{34}+\sg_{1,23}}\\
    &\qquad\qquad\pa{\sg_{45}+\sg_{51}},\notag\\
    \label{eq:PTT6}
    A_{123456}|_{\SD}&=\frac{1}{16}\pa{\sg_{12}+\sg_{23}}\pa{\sg_{34}+\sg_{1,23}}\\
    &~~~~\pa{\sg_{45}+\sg_{1,234}}\pa{\sg_{56}+\sg_{61}}
    .\notag
\end{align}
This suggests the possibility that there may exist a shorter formula for all $n$.

\subsection{General formula}

A conjecture that extends the pattern~\eqref{eq:PTT3}--\eqref{eq:PTT6} to all $n$-particle amplitudes in region $\SD$ is 
\begin{align}
    \label{eq:Simplification}
    A_{1\cdots n}\big|_{\SD}=\frac{1}{2^{n-2}}\prod_{m=2}^{n-1}\pa{\sg_{m,m+1}+\sg_{1,2\cdots m}}.
\end{align}
How sensible is this guess?
Regarding requirements \eqref{eq:Cyclicity}--\eqref{eq:softm}, the region $\SD$ is clearly not cyclically invariant, as it singles out particle 1.
However, we can trivially construct a cyclically invariant answer by using cyclicity to extend~\eqref{eq:Simplification} to other regions $\mathcal{R}_k$ where only particle $k$ has $\omega_k<0$.
The remaining four conditions, when combined with cyclicity, do impose quite nontrivial constraints within $\SD$.
These constraints are all obeyed by our solution \eqref{eq:Simplification}.
In fact, in this form, it is direct to check the soft theorem \eqref{eq:softm} in the last label, and with some more work, in any other label but 1.

The simplified form~\eqref{eq:Recursion} has a simple interpretation: in $\SD$, each factor is $\frac{1}{2}(\pm1\pm1)\in\{-1,0,1\}$, so $A_{1\cdots n}|_{\SD}$ is piecewise-constant and jumps across codimension-one walls where the relevant brackets change sign.
The product form simply makes those walls explicit.

The rest of this work is devoted to proving that the conjecture~\eqref{eq:Simplification} is correct.
The proof uses ideas from time-ordered perturbation theory and proceeds in three parts.
First, one shows that in $\SD$,
\begin{align}
    \label{eq:V0inR1}
    V_{\tlam_2\cdots\tlam_n}\Big|_{\SD}=0,
\end{align}
whilst $\bar{V}_{\tlam_2\cdots\tlam_n}$ remains nonzero.
Second, one must show that with~\eqref{eq:V0inR1}, the recursion~\eqref{eq:Recursion} collapses to become
\begin{align}
    \label{eq:Collapse}
    A_{1\cdots n}\big|_{\SD}=\bar{V}_{\tlam_2\cdots\tlam_n}\Big|_{\SD}.
\end{align}
Finally, one must show that in $\SD$, $\bar{V}_{\tlam_2\cdots\tlam_n}$ reduces to our final formula~\eqref{eq:Simplification}. We briefly sketch each step below.

\subsubsection{Vanishing of \texorpdfstring{$V$}{V}}

We wish to show that within $\SD$,
\begin{align}
    \label{eq:Vanish}
    V_{\tlam_2\cdots\tlam_n}=0
\end{align}
for $n\ge 3$.
This can be interpreted as a causality condition\footnote{It is very reminiscent of the largest-time equation in time-ordered perturbation theory \cite{Caron-Huot:2010fvq}.}
in $\SD$ in the frame in which~\eqref{eq:Pn} holds.
Importantly, all the $\omega$'s appearing in this expression are positive.
This will force at least one of the arguments of the $\Theta$-functions to be negative, implying \eqref{eq:Vanish}. 
For each cut $j$, write left- and right-partial sums as
\begin{align}
    \tlam_L=\sum_{a=2}^j\tlam_a&=\Omega_L(1,\tilde{z}_L),\\
    \tlam_R=\sum_{a=j+1}^n\tlam_a&=\Omega_R(1,\tilde{z}_R),
\end{align}
with $\Omega_{L,R}>0$ and $\tilde{z}_{L,R}$ weighted averages of the $\tilde{z}$'s on each side.
Then
\begin{align}
    \br{\tlam_R\tlam_L}&=\Omega_L\Omega_R\tilde{z}_{R,L},\\
    \br{\tlam_{j+1}\tlam_{j}}&=\omega_j\omega_{j+1}\tilde{z}_{j+1,j}.
\end{align}
Since $\Omega_L\Omega_R>0$ and $\omega_j\omega_{j+1}>0$, the sign of the ratio in the $\Theta$-factor
in \eqref{eq:Vertex} is the sign of $\tilde{z}_{R,L}/\tilde{z}_{j+1,j}$.
A  weighted-variance identity implies that as $j$ runs from $2$ to $n$, there must be at least one cut $j^\star$ for which $\tilde{z}_{R,L}$ has the \emph{same sign} as the adjacent increment $\tilde{z}_{j+1,j}$.
For that $j^\star$, the ratio is positive, so the corresponding factor $\Theta(-\text{positive})=0$, and thus the whole product in~\eqref{eq:Vertex} vanishes.

\subsubsection{Collapsing the recursion}

We have in fact shown something slightly more general: for \emph{every} consecutive $S\subset\{2,\dots,n\}$ with $|S|\ge2$,
\begin{align}
    V_{\tilde\lambda_{S_1}\cdots\tilde\lambda_{S_k}}=0.
\end{align}
Then, from~\eqref{eq:Preamplitudes}, we find that on $\SD$,
\begin{align}
    \label{eq:Abar0}
    \bar{A}_S\Big|_{\SD}=0,
\end{align}
while singletons remain, $\bar{A}_i=1$.
This collapses the recursion.
Using the cyclicity of the color-ordered amplitude, we may write
\begin{align}
    A_{1\cdots n}=A_{2\cdots n1}.
\end{align}
Now apply the recursion \eqref{eq:Recursion} to $A_{2\cdots n1}$, \emph{i.e.}, partition the ordered set
$(2,3,\dots,n)$.
By \eqref{eq:Abar0}, the only nonzero contribution is the all-singleton partition $(2|3|\cdots|n)$, so
\begin{align}
    \label{eq:AtoPTc}
    A_{2\cdots n1}\Big|_{\SD}=-\widehat{\rm PT}_{\tlam_2\tlam_3\cdots\tlam_n}\Big|_{\SD}.
\end{align}
But since $V_{\tlam_2\cdots\tlam_n}=0$, we have $\widehat{\rm PT}=V-\bar{V}=-\bar{V}$ for this list.
Therefore,
\begin{align}
    \label{eq:AequalsVbar}
    A_{1\cdots n}\Big|_{\SD}=A_{2\cdots n1}\Big|_{\SD}=\bar{V}_{\tlam_2\cdots\tlam_n}.
\end{align}
This proves the collapse of the recursion~\eqref{eq:Recursion} to~\eqref{eq:Collapse}.

\subsubsection{Evaluating \texorpdfstring{$\bar{V}_{\tlam_2\cdots\tlam_n}$}{V}}

Last but not least, we reorganize the vertex in terms of sign functions.
Recall that $\bar{V}_{\tlam_2\cdots\tlam_n}$ is defined as
\begin{align}
    \label{eq:VbarStart}
    \bar{V}_{\tlam_2\cdots\tlam_n}=\prod_{m=2}^{n-1}\sg_{m,m+1}\Theta\pa{\frac{\br{\tlam_{2\cdots m}\tlam_{m+1\cdots n}}}{\br{\tlam_m\tlam_{m+1}}}}.
\end{align}
By momentum conservation, $\tlam_{m+1\cdots n}=-\tlam_1-\tlam_{2\cdots m}$, and using the antisymmetry of the bracket, \eqref{eq:VbarStart} becomes
\begin{align}
    \bar{V}_{\tlam_2\cdots\tlam_n}=\prod_{m=2}^{n-1}\sg_{m,m+1}\Theta\pa{\frac{\br{\tlam_1\tlam_{2\cdots m}}}{\br{\tlam_m\tlam_{m+1}}}}.
\label{eq:Vbarafter}
\end{align}
Using the relation between the $\sg$ and $\Theta$ functions, one readily finds that
\begin{align}
    \label{eq:Vbarproduct}
    \bar{V}_{\tlam_2\cdots\tlam_n}=\frac{1}{2^{n-2}}\prod_{m=2}^{n-1}\pa{\sg_{m,m+1}+\sg\pa{\br{\tlam_1\tlam_{2\cdots m}}}}.
\end{align}
Combining \eqref{eq:AequalsVbar} with \eqref{eq:Vbarproduct} recovers exactly our final result \eqref{eq:Simplification}.

\acknowledgments

We are grateful to Nima Arkani-Hamed, Freddy Cachazo, Juan Maldacena, Mark Spradlin and Anastasia Volovich for useful discussions.
AL was supported in part by NSF grant AST-2307888, the NSF CAREER award PHY-2340457, and the Simons Foundation grant SFI-MPS-BH-00012593-09.
AG acknowledges the Roger Dashen membership at the IAS, and additional support from DOE grant DE-SC0009988.
AS \& DS are supported in part by the Simons Collaboration for Celestial Holography.
DS is also supported by the STFC (UK) grant ST/X000664/1.
AS is supported by the Black Hole Initiative and DOE grant DE-SC/0007870, and is grateful for the hospitality of OpenAI where this project was completed. 

\clearpage

\appendix
\onecolumngrid

\section{The master identity}
\label{app:Master}

Consider the well-known identity
\begin{align}
    \label{eq:Identity2}
    \br{\frac{a_1}{(b_2+i\epsilon)}+\frac{a_2}{(b_1+i\epsilon)}}\delta(a_1b_1+a_2b_2)=-\frac{i}{2}\br{\sg(a_1)+\sg(a_2)}\delta(b_1)\delta(b_2),
\end{align}
which readily follows from writing $\frac{1}{b+i\epsilon}=\mathrm{PV}\frac{1}{b}-\frac{i}{2}\delta(b)$, where the $\delta$-function is normalized as in \eqref{eq:DeltaNormalization}.
In this appendix, we use manipulations from time-ordered perturbation theory to derive a powerful generalization of this identity:
\begin{align}
    \label{eq:Master}
    &\phantom{=}\delta\pa{\sum_{k=1}^{n}a_k b_k}\sum_{i=1}^{n}\frac{a_i}{\prod_{j\neq i}(b_j+i\epsilon)}\notag\\
    &=\frac{1}{(2i)^{n-1}}\br{\sum_{i_1}\sg(a_{i_1})+\sum_{i_1<i_2<i_3}\sg(a_{i_1}a_{i_2}a_{i_3})+\sum_{i_1<\cdots<i_{5}}\sg(a_{i_1}a_{i_2}a_{i_3}a_{i_4}a_{i_5})+\ldots}\prod_{i=1}^{n}\delta(b_i).
\end{align}
For example, when $n=3$, this generalizes the identity \eqref{eq:Identity2} to
\begin{align}
    \text{LHS}&=\delta\pa{\sum_{i=1}^{3}a_ib_i}\br{\frac{a_1}{(b_2+i\epsilon)(b_3+i\epsilon)}{+}\frac{a_2}{(b_1+i\epsilon)(b_3+i\epsilon)}{+}\frac{a_3}{(b_1+i\epsilon)(b_2+i\epsilon)}}\\
    =\text{RHS}&=-\frac{1}{4}\br{\sg(a_1)+\sg(a_2)+\sg(a_3)+\sg(a_1)\sg(a_2)\sg(a_3)}\delta(b_1)\delta(b_2)\delta(b_3).
\end{align}
To establish the generalized identity \eqref{eq:Master}, we prove its Fourier transform, that is, the corresponding identity in the time domain where we may think of the $b_i$ as energies.
We then have the following:
\begin{align}
    &\phantom{=}\int\ed^{n}b\,e^{i\sum t_k b_k}\delta\pa{\sum_{k=1}^{n}a_k b_k}\sum_{i=1}^{n}\frac{a_i}{\prod_{j\neq i}(b_j+i\epsilon)}\notag\\
    &=\sum_{i=1}^{n}\int\ed\gamma\int\ed^{n}b\,e^{i\sum(t_k-\gamma a_k)b_k}\frac{a_i}{\prod_{j\neq i}(b_{j}+i\epsilon)}
    =(-2\pi i)^{n-1}\sum_{i=1}^{n}\int_{-\infty}^{\infty}\ed\gamma\,\delta(t_{i}-\gamma a_i)a_i\prod_{j\neq i}\Theta(-t_{j}+\gamma a_{j})\\
    &=(-2\pi i)^{n-1}(2\pi)\int_{-\infty}^{\infty}\ed\gamma\sum_{i=1}^{n}\frac{\pd}{\pd\gamma}\Theta(-t_{i}+\gamma a_i)\prod_{j\neq i}\Theta(-t_{j}+\gamma a_{j})
    =(-2\pi i)^{n-1}(2\pi)\br{\prod_{i=1}^{n}\Theta(-t_{i}+\gamma a_{i})}_{\gamma=-\infty}^{\gamma=\infty}\\
    &=(-2\pi i)^{n-1}(2\pi)\left[\prod_{i=1}^{n}\Theta(a_i)-\prod_{i=1}^{n}\Theta(-a_i)\right].
\end{align}
Finally, Fourier transforming in $t_i$ yields
\begin{align}
   \delta\pa{\sum_{k=1}^{n}a_k b_k}\sum_{i=1}^{n}\frac{a_i}{\prod_{j\neq i}(b_{j}+i\epsilon)}=i^{1-n}\prod_{i=1}^{n}\Theta(a_i)\delta(b_i)-i^{1-n}\prod_{i=1}^{n}\Theta(-a_i)\delta(b_i),
\end{align}
which recovers the master identity after using  $\Theta(x)=\frac{1+\sg(x)}{2}$.

\section{Derivation of recursion relation}
\label{app:Derivation}

This appendix derives the Berends--Giele recursion relation \cite{Berends:1987me} for the off-shell currents in Yang--Mills theory.
In turn, this implies a recursion formula for the planar form factors of the theory.

\subsection{Berends--Giele Recursion}

In QFT, $n$-point scattering amplitudes can be computed from form factors $\mathcal{F}_S$ with one leg off-shell and $n-1$ legs on-shell.
For a color-ordered gluon amplitude, the ordering is inherited from the corresponding form factor.

To avoid notational clutter, we have here taken the last particle to be negative helicity (as opposed to the first one in the main text).
The final on-shell formula is insensitive to this choice.
We thus write
\begin{align}
    \label{eq:AmplitudeFormFactor}
    \mathcal{A}_{1\cdots n}=\lim_{p_n^2\to 0}-ip_n^2\,\mathcal{F}_{1\cdots n-1}\;\delta^{4}\pa{\sum_{i=1}^n p_i},\qquad
    p_n=-\sum_{i=1}^{n-1}p_i.
\end{align}
These planar coefficients satisfy the Berends--Giele (BG) recursion, equivalent to summing over Feynman diagrams with one leg off-shell.
When the rest of the (on-shell) legs are plus-helicity gluons, it is known that this recursion is equal to the one in Self-Dual Yang--Mills theory (SDYM) \cite{Cangemi:1996rx,10.1143/PTPS.123.1}, which reads
\begin{align}
    \label{eq:FormFactor}
    \mathcal F_{1\cdots m}=\frac{1}{p_{1\cdots m}^2+i\epsilon}\sum_{j=1}^{m-1}\br{\tlam_{1\cdots j}\tlam_{j+1\cdots m}}\mathcal{F}_{1\cdots j}\,\mathcal{F}_{j+1\cdots m},
\end{align}
where $p_{1\cdots m}=\sum_{i=1}^m p_i$ and $\tlam_{1\cdots m}=\sum_{i=1}^m\tlam_i$.
This is essentially the equation of motion of the theory and determines its classical solutions.

\subsection{General form factor}

The form factor recursion \eqref{eq:FormFactor} is solved by means of our two-dimensional preamplitudes $\bar{A}_S$ in \eqref{eq:Preamplitudes}, by ``replacing one vertex by PT'' where PT refers to the incomplete Parke-Taylor factor \eqref{eq:PTC}.
Let $(S_1|\cdots|S_A)$ be an ordered partition of the word $(1\cdots m)$, and write the block momenta $K_a=\sum_{i\in S_a}\tlam_{i}$.
We claim the solution to the recursion~\eqref{eq:FormFactor} is
\begin{align}
    \label{eq:FormFactorSolution}
    \mathcal{F}_{1\cdots m}=\sum_{\mathrm{o.p.}}\mathrm{PT}_{K_1\cdots K_A}\;\prod_{a=1}^{A}\Big(\bar A_{S_a}\,\delta_{S_a}\Big),
\end{align}
where the sum is over all possible ordered partitions of $(1\cdots m)$ into $A$ blocks $S_a$, as well as over the number $A=1,\ldots,m$ of blocks. As in the main text, $\delta_{S_a}$ denotes the product of $\delta(z_{i,i+1})$ internal to the block $S_a$.

To see this, we first establish \eqref{eq:PT}, which we quote again here for the benefit of the reader: 
\begin{align}
    \label{eq:PT2}
    \mathrm{PT}_{1\cdots n}-\delta_{1\cdots n}\,V_{\tlam_1\cdots\tlam_n}=\sum_{j=1}^{n-1}\frac{\br{\tlam_{1\cdots j}\tlam_{j+1\cdots n}}}{p_{1\cdots n}^2+i\epsilon}\mathrm{PT}_{1\cdots j}\,\mathrm{PT}_{j+1\cdots n}.
\end{align}
Away from the half-collinear regime, we have $\delta_{1\cdots n}=0$, the $i\epsilon$'s on the RHS may be neglected, and this identity is a standard algebraic identity for Parke-Taylor factors.
The $i\epsilon$'s are important in the half-collinear limit and the general identity~\eqref{eq:PT2} follows from~\eqref{eq:Master} with the values
\begin{align}
    a_r=-\br{\tlam_{1\cdots r}\tlam_{r+1\cdots n}}\prod_{\ell\neq r}\br{\tlam_\ell\tlam_{\ell+1}},\qquad
    b_r=p_{r,r+1}^2,
\end{align}
for $r=1,\ldots,n-1$, while $a_n=\prod_{\ell}[\ell,\ell+1]$ and $b_n=p_{1\ldots n}^2$.

Next, we will verify \eqref{eq:FormFactorSolution} is a solution by inserting it into the RHS of \eqref{eq:FormFactor}. For each term in the sum over $j$ in \eqref{eq:FormFactor}, we now have sums over all ordered partitions $(1\cdots j)=(L_1|\cdots|L_B)$ and $(j+1\cdots m)=(R_1|\cdots|R_C)$:
\begin{align}
    \mathcal{F}_{1\cdots m}&=\sum_{j=1}^{m-1}\sum_{\,\text{o.p.}}\frac{\br{\tlam_{1\cdots j}\,\tlam_{j+1\cdots m}}}{p^2_{1\cdots m}+i\epsilon}\,\mathrm{PT}_{L_1L_2\cdots L_{B}}\mathrm{PT}_{R_1R_2\cdots R_{C}}\notag\\
    &\qquad\qquad\qquad\qquad\times\bar{A}_{L_1}\cdots\bar{A}_{L_{B}}\bar{A}_{R_1}\cdots\bar{A}_{R_{C}}\delta_{L_1}\cdots\delta_{L_{B}}\delta_{R_1}\cdots\delta_{R_{C}}\,,
\end{align}
where we have abused notation slightly, denoting both the set and its block momentum by the same letter $L_a$ or $R_b$.
Note that $\tilde{\lambda}_{1\cdots j}=\sum_k\tilde{\lambda}_{L_k}$ and $\tilde{\lambda}_{j+1\cdots m}=\sum_k\tilde{\lambda}_{R_k}$.
The second line depends only on the total partition $(S_1|\cdots|S_A)=(L_1|\cdots|L_B|R_1|\cdots|R_C)$, with $A=B+C$, and does not know about the separation into ``left'' and ``right''.
For a fixed such partition, the $j$-sum is exactly the PT recursion \eqref{eq:PT2}: it produces (i) a PT for the combined partition, and (ii) a contact term $\delta\,V$.

In equations, we have
\begin{align}
    \mathcal{F}_{1\cdots m}&=\sum_{\text{n.t.p.}}\bar{A}_{S_1}\cdots\bar{A}_{S_{A}}\delta_{S_1}\cdots\delta_{S_{A}}
    \pa{\sum_{B+C=A}\frac{\br{\tlam_{1\cdots j}\,\tlam_{j+1\cdots m}}}{p^2_{1\cdots m}+i\epsilon}\,\mathrm{PT}_{L_1\cdots L_{B}}\mathrm{PT}_{R_{1}\cdots R_{C}}}\\
    &=\sum_{\text{n.t.p.}}\bar{A}_{S_1}\cdots\bar{A}_{S_{A}}\delta_{S_1}\cdots\delta_{S_{A}}\mathrm{PT}_{S_1S_2\cdots S_{A}}-\delta_{1\cdots m}\sum_{\text{n.t.p.}}\bar{A}_{S_1}\cdots\bar{A}_{S_{A}}V_{S_1S_2\cdots S_{A}}\\
    &=\sum_{\text{n.t.p.}}\bar{A}_{S_1}\cdots\bar{A}_{S_{A}}\delta_{S_1}\cdots\delta_{S_{A}}\mathrm{PT}_{S_1S_2\cdots S_{A}}+\delta_{1\cdots m}\bar{A}_{1\cdots m},
\end{align}
where n.t.p. denotes nontrivial partitions (having more than one block).
The PT piece reproduces the nontrivial $A$-block contribution on the LHS of \eqref{eq:FormFactorSolution}.
The contact piece $\delta\,V$ precisely stitches blocks together by a $V$-vertex, and the resulting sum over nontrivial partitions is the recursion~\eqref{eq:Preamplitudes} for the preamplitudes.
The final term in the third line is the missing 1-term partition, thus recovering \eqref{eq:FormFactorSolution}.

\subsection{LSZ Reduction}

The single-minus amplitude is obtained by putting the ``last'' leg on-shell, as in \eqref{eq:AmplitudeFormFactor}.
On the support of the collinear $\delta$-functions inside each block of the form factor \eqref{eq:FormFactorSolution}, every block momentum $K_a$ is null.
The only remaining singular factor is the PT term associated to adjacent channels.
The on-shell limit of a PT factor is evaluated using the master identity in App.~\ref{app:Master}, resulting in
\begin{align}
    \label{eq:LSZ}
    \lim_{p_n^2\to0}p_n^2\,\mathrm{PT}_{K_1\cdots K_k}\,\delta^{4}\pa{\sum_{a=1}^k K_a + p_n}=\widehat{\mathrm{PT}}_{K_1\cdots K_k}\delta_{1\cdots k,n}\delta^2\pa{\sum_j\tilde{\lambda}_j}
    =(V_{K_1\cdots K_k}-\bar{V}_{K_1\cdots K_k})\delta_{1\cdots k,n}\delta^2\pa{\sum_j\tilde{\lambda}_j}.
\end{align}
After stripping off the universal momentum-conservation support (here, $\sum_{i=1}^n\tlam_i=0$, so $\tlam_n=-\sum_{i=1}^{n-1}\tlam_i$), \eqref{eq:FormFactorSolution} and \eqref{eq:LSZ} yield, after some algebra, the final result \eqref{eq:Recursion}:
\begin{align}
    A_{1\cdots n}=-\sum_{\substack{(1\cdots n-1)=S_1|\dots|S_k\\ k\ge 1}}\widehat{\mathrm{PT}}_{\tlam_{S_1}\cdots\tlam_{S_k}}\;\prod_{a=1}^{k}\bar A_{S_a}\,.
\end{align}

\bibliographystyle{utphys}
\bibliography{SMGA.bib}

\end{document}